\documentclass[12pt]{iopart}
\usepackage{iopams,mathrsfs,graphicx,hyperref,color}
\usepackage[square,numbers]{natbib}
\usepackage{ulem}
\usepackage{mathbbol}

\begin{document}
\title[Biased random walk on a random comb in presence of stochastic resetting]{Biased random walk on random networks in presence of stochastic resetting: Exact results}
\author{Mrinal Sarkar$^{1,2}$ and
Shamik Gupta$^{3}$}
\address{$^1$Department of Physics, Indian Institute of Technology Madras, Chennai 600036, India \\$^2$Institute for Theoretical Physics, University of Heidelberg,
	Philosophenweg 19, D-69120 Heidelberg, Germany\\
$^2$Department of Theoretical Physics,  Tata Institute of Fundamental Research,  Homi Bhabha Road, Mumbai 400005, India}
\ead{$^1$mrinal@physics.iitm.ac.in, $^3$shamikg1@gmail.com}
\vspace{10pt}

\begin{abstract}
We consider biased random walks on random networks constituted by a random comb comprising a backbone with quenched-disordered random-length branches.  The backbone and the branches run in the direction of the bias.  For the bare model as also when the model is subject to stochastic resetting, whereby the walkers on the branches reset with a constant rate to the respective backbone sites, we obtain exact stationary-state static and dynamic properties for a given disorder realization of branch lengths sampled following an arbitrary distribution.  We derive a criterion to observe in the stationary state a non-zero drift velocity along the backbone.  For the bare model, we discuss the occurrence of a drift velocity that is non-monotonic as a function of the bias,  becoming zero beyond a threshold bias because of walkers trapped at very long branches. Further, we show that resetting allows the system to escape trapping, resulting in a drift velocity that is finite at any bias.
\end{abstract}
\maketitle

%

Random walk (RW) on random networks such as random comb (RC) lattices, inspired by Pierre de Gennes' `\textit{Ant-in-a-Labyrinth}'~\cite{de1976percolation}, is a much-studied research topic~\cite{barma1983directed, white1984field, dhar1984diffusion, bunde1986diffusion, goldhirsch1987biased, havlin1987diffusion, havlin1987anomalous, aslangul1994analytic, balakrishnan1995transport, pottier1995diffusion, mitran2013biased, yuste2016anomalous, demaerel2018death, hart2020compact, kotak2022bias}. An RC, comprising a backbone with random-length branches, encodes essential features of physical problems, e.g., finitely-ramified fractals and percolation clusters~\cite{stauffer1979scaling,rammal1983random,sahimi1993flow}.  Biased RW on RCs yields many nontrivial results, e.g., a drift varying non-monotonically with bias~\cite{barma1983directed, white1984field, dhar1984diffusion}, anomalous diffusion~\cite{havlin1987anomalous, pottier1995diffusion, bunde1986diffusion, balakrishnan1995transport}. Dynamics on comb-like structures finds wide applications in modelling many natural phenomena, e.g., transport in spiny dendrites~\cite{mendez2013comb}, rectification in biological ion channels~\cite{cecchi1996negative}, superdiffusion of ultra-cold atoms~\cite{iomin2012superdiffusive}, reaction-diffusion processes~\cite{agliari2014slow}, crowded-environment diffusion~\cite{benichou2015diffusion}, cancer proliferation~\cite{iomin2006toy}, and even human migration along river networks~\cite{campos2006transport}.

In recent years, stochastic resetting has been extensively studied in the area of nonequilibrium statistical mechanics. The setup involves repeated interruptions of a dynamics at random times with a reset to the initial condition~\cite{r0,evans2020stochastic}.  Resetting results in a nonequilibrium stationary state (NESS) with remarkable static and dynamic features. Examples include a wide spectrum of dynamics: diffusion~\cite{r7, nagar2016diffusion, majumdar2018spectral, den2019properties, r17, masoliver2019telegraphic, ray2020diffusion}, random walks~\cite{montero2016directed,r20}, L\'{e}vy flights~\cite{kusmierz2014first},  Bernoulli trials~\cite{belan2018restart}, discrete-time resets~\cite{coghi2020large}, active motion~\cite{kumar2020active} and transport in cells~\cite{bressloff2020modeling},  search problems~\cite{evans2013optimal, r3, pal2017first, falcon2017localization, chechkin2018random, r12, r21},  RNA-polymerase dynamics~\cite{r25, r26},  enzymatic reactions~\cite{reuveni2016optimal}, ecology~\cite{boyer2014random,giuggioli2019comparison}, interacting systems~\cite{r23,r24,durang2014statistical,ising-resetting,sarkar2022synchronization,basu2019symmetric, karthika2020totally}, stochastic thermodynamics~\cite{fuchs2016stochastic}, quantum dynamics~\cite{r4},  etc.

\begin{figure*}[!ht]
	\centering
	\includegraphics[scale=0.62]{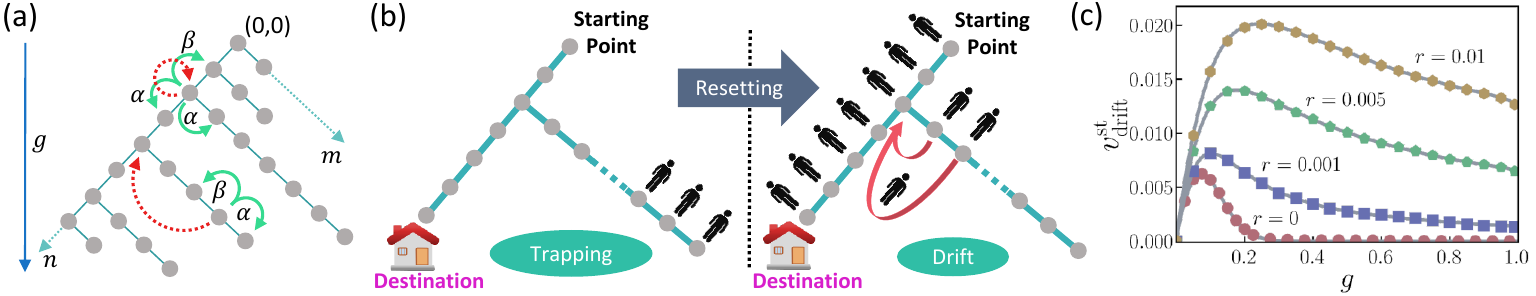}
	\caption{(a) A random comb comprising a backbone, with random-length branches; Broken and continuous arrows denote respectively resetting and biased hopping in presence of bias $g$.  (b) Dramatic consequence of resetting on stationary-state transport shown schematically: No resetting results in walkers trapped towards the end of very-long branches (shown here is one such branch) and consequently, zero drift velocity along backbone. Long-range instantaneous jumps due to resetting allow walkers from the open end to get to the backbone, implying no trapping (thus, vanishing probability to find the walkers towards the open end) and hence, a nonzero drift velocity along backbone. (c) Stationary-state drift velocity versus $g$ from theory (Eq.~(\ref{eq:drift_vel}),  continuous line) and numerics (symbols), with $W=0.5$, number of backbone sites $N=200$ and for a typical branch-length realization sampled from exponential distribution~(\ref{eq:exponential-powerlaw-PL}) with $M=20$ and $\xi=5$. {Numerics correspond to standard Monte Carlo simulations of the dynamics~\cite{SM}.}}
	\label{fig:drift_vel_numerics}
\end{figure*}

In this Letter,  we revisit the classic problem of biased non-interacting RWs in continuous time and on RC, with a twist, namely, with stochastic resets.  As regards resetting, we address an unexplored theme: resetting in a system with quenched disorder.  The RC-backbone (Fig.~\ref{fig:drift_vel_numerics}(a)) is a one-dimensional ($1d$) lattice of $N$ sites, to each of which is attached a branch of a $1d$ lattice with a random number of sites (all lattice-spacings are unity).  Let $M$ denote the maximum-allowed branch length. Denote the sites by ($n,m$), wherein $0\leq n \leq N-1$ labels the backbone sites and $0 \leq m \leq L_{n}$ labels the $(L_n+1)$ number of sites on the branch attached to the $n$-th backbone site.  The site ($n,m=0$) being shared by the backbone and the branch,  we will from now on refer to branch sites as those with $m>0$. The $L_{n}$'s are quenched-disordered random variables drawn independently from an arbitrary distribution $\mathcal{P}_L$.  The backbone and branches run along a field or a bias with strength $g$; $0 < g < 1$.  Representative $\mathcal{P}_L$'s are an exponential and a power-law given respectively by
\begin{eqnarray}
	\mathcal{P}_L= \left\{ \begin{array}{lr} \frac{{1 - e^{- {1}/{\xi}}} }{1-e^{- (M+1)/\xi}} e^{- L/\xi};~0 \le L \le M,
		\\ \left[ \sum_{L=1}^{M} L^{-k} \right]^{-1}L^{-k} ;~k>1,~1 \le L \le M.
	\end{array}
	\right.
	\label{eq:exponential-powerlaw-PL}
\end{eqnarray}
As $M \to \infty$,  power-law $\mathcal{P}_L$ has finite mean for $k>2$, while that of the exponential is always finite. The dynamics in time $[t,t+{\rm d}t]$ involves a walker on a site performing either (i) biased hopping with probability $1-r{\rm d}t$: hop to nearest-neighbor (NN) site(s) along (respectively,  against) the bias with rate $\alpha \equiv W(1 + g)$ (respectively, $\beta \equiv W(1-g)$), or,  (ii) resetting with probability $r{\rm d}t$. The latter involves (a) reset from a branch to the respective backbone site; (b) reset from a backbone site to itself, with $r$ the resetting rate.  We assume respectively periodic and reflecting boundary conditions for backbone and open end of the branches, and define $f \equiv \alpha/\beta >1$.  

The system,  in absence ($r=0$) and presence ($r\ne 0$) of resetting, settles at long times into an NESS. Even with $r=0$, analytical characterization of the NESS is a long-standing open problem, with approximate analysis pursued until now. For instance, in analyzing transport properties,  physical arguments assuming zero current in the branches~\cite{barma1983directed, white1984field}, or, a mean-field approach~\cite{havlin1987anomalous, balakrishnan1995transport, pottier1995diffusion} based on self-consistent scaling and continuous-time random walk was invoked. A remarkable revelation is that, for exponential $\mathcal{P}_L$, the stationary-state drift velocity along the backbone, $v^{\rm st}_{\rm {drift}}$,  varies non-monotonically with $g$ as $M \to \infty$, becoming zero beyond a threshold $g_c$ because of trapping at long branches. 

We motivate our study thus: Referring to Fig. ~\ref{fig:drift_vel_numerics}(b), consider a random walker aiming to reach a destination lying ahead (which defines the bias direction) on the backbone but is unaware of the path to it.  At every branch-backbone junction,  it either enters the branch or continues on the backbone. While on a branch, it may at a random time realize that it may not eventually get to the destination, and deterministically walks back to the junction point.  The deterministic motion being on a fast time scale compared to the RW-dynamics may be treated as an instantaneous resetting on the scale of the latter.  A drift along the backbone at long times implies that the walker eventually reaches the destination.

Here,  we report for biased RWs on RC \textit{exact} NESS static and dynamic properties both in absence and presence of resetting and for \textit{any} disorder realization $\{L_n\}$ corresponding to \textit{arbitrary} $\mathcal{P}_L$. The NESS-distribution of walkers (Eq. ~(\ref{eq:p_st_with_r_text})) and the associated $v^{\rm st}_{\rm {drift}}$ (Eq.~(\ref{eq:drift_vel})) hold for general $N$; for the latter, we validate earlier results obtained using approximations as $N \to \infty$ and for exponential $\mathcal{P}_L$~\cite{white1984field}.  Further, we propose and verify a criterion (Eq.~(\ref{eq:trapping_condition})), valid for arbitrary $\mathcal{P}_L$, to observe trapping and hence a vanishing drift velocity.  We establish a dramatic consequence of resetting~(Fig.~\ref{fig:drift_vel_numerics}(b)): In its absence,  a choice of $\mathcal{P}_L$ that leads to trapping of walkers towards the open end of long branches and a vanishing drift velocity results, with resetting,  in a nonzero drift velocity. Resetting allows walkers to make long-range instantaneous jumps to reach the backbone from the open end, implying no trapping and consequently,  nonzero drift velocity. This Letter reports a rare example of a system with quenched disorder for which we obtain the \textit{exact} NESS (i) in absence and presence of resetting,  (ii) for any disorder realization,  and (iii) in the thermodynamic limit ($N \to \infty$) as well as for finite $N$. 

Resetting on comb-like structures was invoked in discussing diffusion process in three dimensions~\cite{domazetoski2020stochastic}, random walks on comb graphs with equal-length side-chains~\cite{gonzalez2021diffusive}, and diffusion in a two-dimensional comb with continuously-distributed branches~\cite{singh2021backbone}.  Our setup involving combs with random branch-lengths and focus on exact NESS deviate markedly from these studies.

To proceed,  define $P_{n,m}(t)\equiv P(n,m,t|0,0,0)$ as the conditional probability for a walker to be on site $(n,m)$ at time $t>0$,  given that it was on $(0,0)$ at $t=0$. With normalization $\sum_{n=0}^{N-1} \sum_{m=0}^{L_n} P_{n,m}(t)=1$, $P_{n,m}(t)$ satisfies the master equation (ME):
\begin{eqnarray}
	\dot{P}_{n,m}= \mathcal{L} P_{n,m}(t) - rP_{n,m}(t) + r\delta_{m,0} \sum_{m'=0}^{L_{n}} P_{n,m'}(t),
	\label{eq:ME}
\end{eqnarray}
with dot denoting time derivative. With $\mathcal{W}_{(n',m') \to (n,m)}$ the transition rate from $(n',m')$ to $(n,m)$ and sum running over all $(n',m')$ that are NN-sites of $(n,m)$,  the term $\mathcal{L} P_{n,m}(t) \equiv \sum\limits_{(n',m')} \left[\mathcal{W}_{(n',m') \to (n,m)} P_{n',m'}(t) - \mathcal{W}_{(n,m) \to (n',m')} P_{n,m}(t) \right]$ represents ways in which $P_{n,m}(t)$ changes due to biased-RW dynamics. The second and third terms on the right hand side (rhs) of Eq.~(\ref{eq:ME}) stand for resetting.  The former represents gain in probability at the backbone site due to resetting,  while the latter denotes the corresponding loss in probability.  

To solve (\ref{eq:ME}) for $P_{n,m}(t)$'s for a given realization $\{L_n\}$,  apply Laplace transformation (LT) to Eq.~(\ref{eq:ME}): $\widetilde{P}_{n,m}(s) \equiv \int_{0}^{\infty} {\rm d}t~e^{-st} P_{n,m}(t)$~\cite{aslangul1994analytic}.  The ME for branch sites,  $\dot{P}_{n,m}(t)=  \alpha P_{n, m-1}(t) - (\beta + r) P_{n,m}(t) +  (1- \delta_{m, L_{n}}) \left[ \beta P_{n,m+1}(t) - \alpha P_{n,m}(t) \right]$, involves three sites, except for the reflecting end ($m=L_{n}$) that involves the last two branch sites.  Applying LT to the ME for $m=L_n$ gives $\widetilde{P}_{n, L_{n}-1}(s) = \left((s + \beta + r)/\alpha\right)\widetilde{P}_{n,L_{n}}(s)$. 
This helps to relate the LT-transformed probabilities on two consecutive branch sites by considering successively the LT-transformed branch-ME for $m = L_{n}-1,  \dots, 1$.  We get~\cite{SM}: $\widetilde{P}_{n,m}(s) = \Gamma_{L_{n} -m +1}  \widetilde{P}_{n, m-1} (s);~~m=1,\dots, L_{n},
\label{eq:P_nm_recursion_rel}
$
with finite continued fraction $\Gamma_\mathcal{M}(s, r)$ being
\begin{eqnarray}
	\Gamma_\mathcal{M} \equiv \frac{1}{ \frac{s + \alpha + \beta + r}{\alpha} - \frac{\beta}{\alpha}     \frac{1}{  \frac{s + \alpha+ \beta + r}{\alpha} - \frac{\beta}{\alpha}     \frac{1}{ \ddots \frac{s + \alpha+ \beta + r}{\alpha} - \frac{\beta}{\alpha} \frac{1}{\frac{s + \beta + r}{\alpha} } } }},
	\label{eq:Lambda_m_def_text}
\end{eqnarray}
containing $\mathcal{M}$ terms in the denominator. In particular,  $\widetilde{P}_{n,1}(s) = \Gamma_{L_{n}}(s, r) \widetilde{P}_{n,0}(s)$. A remarkable transformation  $\cosh \theta \equiv \sqrt{f} \left((s+ \alpha + \beta + r)/(2\alpha)\right)=(2W + r)/(2W \sqrt{1-g^{2}}) \left(1 + s/(2W +r)\right)$ evaluates $\Gamma_\mathcal{M}$ in closed form, yielding for $\mathcal{M} = L_n$,
\begin{eqnarray}
	\Gamma_{L_{n}} =  \sqrt{f} \frac{\sinh L_{n} \theta - \sqrt{f} \sinh (L_{n} -1) \theta}{\sinh (L_{n} +1) \theta - \sqrt{f} \sinh L_{n} \theta}.
	\label{eq:Omega_n_closed_form_text}
\end{eqnarray}
The recursion $\widetilde{P}_{n,m}(s) = \Gamma_{L_{n} -m +1}  \widetilde{P}_{n, m-1} (s)$ and the closed-form $\Gamma_\mathcal{M}$ give
\begin{eqnarray}
	\frac{\widetilde{P}_{n,m}(s)}{\widetilde{P}_{n,0} (s)}=f^{m/2} \frac{\sinh (L_{n} -m +1) \theta - \sqrt{f} \sinh (L_{n} -m) \theta}{\sinh (L_{n} + 1) \theta - \sqrt{f} \sinh L_{n} \theta}.
	\label{eq:prob_Laplace_branch}
\end{eqnarray}

We now apply LT to the ME for the backbone: 
\begin{eqnarray}
\dot{P}_{n,0}(t)
&=&  \alpha [ (1-\delta_{n,0}) P_{n-1,0}(t)+ \delta_{n,0} P_{N-1,0}(t)] \nonumber \\
&&+ \beta [ (1-\delta_{n,N-1})P_{n+1,0}(t)+ \delta_{n,N-1} P_{0,0}(t)]+ \beta P_{n,1}(t) \nonumber \\
&& - (2\alpha + \beta) P_{n,0}(t) + r \sum_{m'=1}^{L_{n}} P_{n,m'}(t);~~0 \leq n \leq N-1,
\end{eqnarray}
where effects of resetting from backbone sites onto themselves cancel out.  We get 
\begin{eqnarray}
s \widetilde{P}_{n,0}(s) - \delta_{n,0}&=&\alpha [ (1-\delta_{n,0}) \widetilde{P}_{n-1,0}(s) + \delta_{n,0} \widetilde{P}_{N-1,0}(s) ]\nonumber \\
&&  + \beta [ (1-\delta_{n,N-1}) \widetilde{P}_{n+1,0}(s) + \delta_{n,N-1} \widetilde{P}_{0,0}(s)  ] + \beta \widetilde{P}_{n,1}(s)\nonumber \\
&& -(2\alpha + \beta) \widetilde{P}_{n,0}(s) + r \sum_{m'=1}^{L_n} \widetilde{P}_{n,m'} (s).
\end{eqnarray}
  For each $n$,  this ME involves three consecutive backbone sites and all the attached branch sites.  Using $\widetilde{P}_{n,1}(s) = \Gamma_{L_{n}} \widetilde{P}_{n,0}(s)$ and defining $\Delta_{L_{n}}(s, r)$ as $\Delta_{L_{n}} \widetilde{P}_{n,0} (s) \equiv \sum_{m'=1}^{L_n} \widetilde{P}_{n,m'} (s)~\forall~n,\label{eq:Deltan_defn}$ replace the LT-transformed branch-site probabilities in the ME with $\widetilde{P}_{n,0}(s)$, giving 
\begin{eqnarray}  
  s \widetilde{P}_{n,0}(s) - \delta_{n,0} &=& \alpha [ (1-\delta_{n,0}) \widetilde{P}_{n-1,0}(s) + \delta_{n,0} \widetilde{P}_{N-1,0}(s) ]\nonumber \\
  &&  + \beta [ (1-\delta_{n,N-1}) \widetilde{P}_{n+1,0}(s) + \delta_{n,N-1} \widetilde{P}_{0,0}(s)  ] + \beta \Gamma_{L_{n}} \widetilde{P}_{n,0}(s) \nonumber \\
  &&-(2\alpha + \beta) \widetilde{P}_{n,0}(s) + r \Delta_{L_{n}} \widetilde{P}_{n,0}(s).
  \end{eqnarray} These $N$ coupled linear equations involving only the backbone sites write as a matrix equation: 
\begin{eqnarray}
	\textbf{A}\widetilde{\textbf{P}}(s) = \textbf{E},
	\label{eq:matrix_eqn_text}
\end{eqnarray}
with $\widetilde{\textbf{P}}(s) \equiv \left(\widetilde{P}_{0,0}(s), \widetilde{P}_{1,0}(s), \dots, \widetilde{P}_{N-1,0}(s)\right)^T$, ~$\textbf{E} \equiv \left( 1,0,\dots,0 \right)^T$,  $T$ denoting transpose. The matrix $\textbf{A}$ has elements $A_{n,n'} = - \alpha \delta_{n-1,n'} + C_{n} \delta_{n,n'} - \beta \delta_{n+1,n'}$ for $ 0\leq n,n'\leq N-1,$ with $\delta_{-1,n'}= \delta_{N-1,n'}$,  $\delta_{N,n'}= \delta_{0,n'}$, $C_n \equiv s + 2\alpha + \beta(1 - \Gamma_{L_{n}}) - r\Delta_{L_{n}}$,
where $\Delta_{L_{n}}$ on using Eq.~(\ref{eq:prob_Laplace_branch}) evaluates as~\cite{SM}: $\Delta_{L_{n}} = (\beta/(s+r)) \left( f - \Gamma_{L_{n}} \right)$.  
Equation~(\ref{eq:matrix_eqn_text}) gives $\widetilde{\textbf{P}}(s) = \textbf{A}^{-1}\textbf{E}$,  which evaluated numerically yields LT-transformed backbone-site probabilities for a given realization $\{L_n\}$; the same for branch sites are given by Eq.~(\ref{eq:prob_Laplace_branch}).  Inverse LT of $\widetilde{P}_{n,m}(s)$'s so obtained yields $P_{n,m}(t)~\forall~n,m,t>0$.

We are interested in the transport properties in the NESS. The latter is characterized by time-independent probabilities $P_{n,m}^{\rm{st}} = \lim_{t \to \infty } P_{n,m} (t)$,  obtained from Eq. ~(\ref{eq:matrix_eqn_text}) by using the final value theorem (FVT): $P_{n,m}^{\rm{st}}= \lim_{s \to 0} s\widetilde{P}_{n,m}(s)$. Consider $s \to 0$ such that for any $g$ and $r>0$, $s/g \ll 1$ and $s/r \ll 1$.  One then obtains from Eq.~(\ref{eq:Omega_n_closed_form_text}) that $\Gamma_{L_{n}} (s,r>0)|_{s \to 0} = \Lambda_{1,L_{n}}/ \Lambda_{0,L_{n}}$,  with $\Lambda_{m,L_{n}} \equiv (f^{m/2}/2) [  \lambda^{L_{n}-m} \left( \lambda - \sqrt{f} \right) -  \lambda^{-L_{n}+m} \left( 1/\lambda- \sqrt{f} \right) ]$; $m=0,1,\dots, L_n$,  and $\lambda \equiv (2W + r)/(2W \sqrt{1-g^{2}})  [1+ \sqrt{1- (4 W^{2} (1 -g^{2}))/((2W +r)^{2})} ]$, while $\Delta_{L_{n}}(s,r>0)|_{s \to 0} = (\beta/r) \left( f - \Gamma_{L_{n}}(s,r>0)|_{s \to 0}\right)$.  We thus get $C_{n}|_{s \to 0} = s + 2 \alpha + \beta \left( 1- \Gamma_{L_{n}}(s,r>0)|_{s \to 0} \right) - r \Delta_{L_{n}}(s,r>0)|_{s \to 0}= \alpha + \beta$. Equation~(\ref{eq:matrix_eqn_text}), on applying FVT, thus gives stationary-state backbone-ME: $ C_{n}|_{s \to 0} P_{n,0}^{\rm{st}} = \alpha [ (1-\delta_{n,0}) P_{n-1,0}^{\rm{st}} + \delta_{n,0} P_{N-1,0}^{\rm{st}}] + \beta [ (1-\delta_{n,N-1}) P_{n+1,0}^{\rm{st}} + \delta_{n,N-1} P_{0,0}^{\rm{st}} ]$; the rhs denotes gain in probability,  which is balanced by the left denoting the corresponding probability loss. $C_{n}|_{s \to 0}$ then gives stationary-state transition rate out of the $n$-th backbone site.  

The result $C_{n}|_{s \to 0}=\alpha+\beta$ is non-trivial and interesting: it (i) does not involve $r$,  (ii) is independent of $n$, or, equivalently, $L_{n}$,  (iii) has the same value as for $L_n =0$  (for $L_n=0$,  $\Gamma_{L_{n}}=f$ and $\Delta_{L_{n}}=0$ give $C_n|_{s\to 0}=[s+2\alpha+\beta(1-f)]|_{s\to 0}=\alpha+\beta$.). Remarkably, the stationary-state backbone-ME has no branch-effects although the underlying dynamics involves hopping and resetting and includes backbone and branch sites. Indeed, this ME is mathematically equivalent to that for single-site probabilities $p_n^{\rm st};~n=0,1,\ldots,N-1$ for non-interacting random walkers undergoing only hopping to NN sites with rates $\alpha$ and $\beta$ on a $1d$ periodic lattice of $N$ sites.  This equivalence holds key to our exact results on $v^{\rm st}_{\rm {drift}}$.

The aforementioned equivalence is by no means obvious and holds only in the NESS.  Then, if the stationary-state backbone-ME in presence of hopping and resetting is the same as the one on a $1d$ periodic lattice with only hopping, how do branch-effects manifest in the former? The answer lies in the normalization of the stationary-state probabilities.  The stationary-state ME yields in both cases a uniform probability: uniform ($P_{n,0}^{\rm st}=P^{\rm st}~\forall~n$) over the backbone,  uniform ($=p^{\rm st}$) over the $1d$ periodic lattice.  The normalization condition however reads differently: $\sum_{n=0}^{N-1} \sum_{m=0}^{L_n} P_{n,m}^{\rm{st}} = 1$ and $\sum_{n=0}^{N-1} p^{\rm{st}} = 1$.  Note that for RC, the branch-site probabilities are not uniform.  Applying FVT to the equation defining $\Delta_{L_{n}}$ gives $\Delta_{L_{n}} (s,r>0)|_{s \to 0}P^{\rm st}=(\beta/r) \left( f - \Gamma_{L_{n}}(s,r>0)|_{s \to 0}\right)P^{\rm st}=\sum_{m'=1}^{L_n} P_{n,m'}^{\rm st}$, which used in the normalization condition gives $P^{\rm{st}} =  (1/N)  {\left[1 + (1/N) \sum_{n=0}^{N-1} \Delta_{L_{n}}(s,r>0)|_{s \to 0} \right]}^{-1}$, while $p^{\rm st}=1/N$. The stationary-state branch-site probabilities are obtained by applying FVT to Eq.~(\ref{eq:prob_Laplace_branch}), yielding $P_{n,m}^{\rm{st}} = (\Lambda_{m,L_{n}}/ \Lambda_{0,L_{n}}) P^{\rm{st}}$.

To obtain the NESS for no-resetting case, we first set $r=0$ and consider $s \to 0$ such that $s/g \ll 1$ for any $g$, to get $\Gamma_{L_{n}} (s,0)|_{s \to 0} = f$ and $\Delta_{L_{n}}(s,0)|_{s \to 0} = \left( {f}/({f-1})\right)  \left( f^{L_{n}} -1\right)$,  yielding $C_{n}|_{s \to 0} = \alpha + \beta$.  Using the equivalence of the stationary-state backbone-ME with that for a $1d$ periodic lattice and the following steps as invoked above for $r\ne0$ yield the exact expression for the backbone-site probabilities for a given realization $\{ L_{n}\}$ as $P^{\rm{st}} =  (1/N)  {\left[1 + (1/N)  \sum_{n=0}^{N-1} \Delta_{L_{n}}(s,r=0)|_{s \to 0} \right]}^{-1}$;
the same for the branch sites are given by $P_{n,m}^{\rm st}=f^m P^{\rm st}$.  
We thus obtain \textit{exact} stationary-state probabilities on \textit{all RC-sites} both in presence and absence of resetting and for a given realization $\{ L_{n}\}$,  one of our key results applicable to any RC as in Fig.~\ref{fig:drift_vel_numerics}(a). The backbone probability has the form
\begin{eqnarray}
	P^{\rm{st}} =  \frac{1}{N}  \frac{1}{\left[1 + \frac{1}{N}  \sum_{n=0}^{N-1} \Delta_{L_{n}}(s,r)|_{s \to 0} \right]},
	\label{eq:p_st_with_r_text}
\end{eqnarray}
with 
\begin{eqnarray}
	\Delta_{L_{n}}(s,r)|_{s \to 0} = \left\{ \begin{array}{lr} \frac{f}{f-1}  \left( f^{L_{n}} -1\right) ;&~r = 0,\\
		\frac{\beta}{r} \left( f - \Gamma_{L_{n}}(s,r>0)|_{s \to 0}\right) ;&~r \neq 0.
	\end{array}
	\right.
	\label{eq:Delta_Ln}
\end{eqnarray}

To compute $ v^{\rm st}_{\rm {drift}}$, consider the equivalent $1d$ system of non-interacting walkers.  The probability $p_n(t)$ to be on site $n$ at time $t$ while starting from $n=0$ at $t=0$ satisfies the ME $\dot{p}_{n}(t)= \alpha [ (1-\delta_{n,0}) p_{n-1}(t) + \delta_{n,0} p_{N-1}(t) ] + \beta [ (1-\delta_{n,N-1}) p_{n+1}(t) + \delta_{n,N-1} p_{0}(t) ] - (\alpha + \beta) p_{n}(t)$.  Let {$\mathbb{p}_{n + l_{n}N}(t)$} be the probability that a walker starting from $n=0$ at $t=0$ and undergoing integer $l_n \in (-\infty,\infty)$ number of  turns round the periodic lattice arrives at site $n$ at time $t$.  Evidently, $p_{n}(t) = \sum_{l_{n}} {\mathbb{p}_{n + l_{n}N}(t)}~\forall~n,t$, and {$\mathbb{p}_{n + l_{n}N}(t)$} satisfies the same ME as $p_{n}(t)$. The average displacement in time $t$ is $\langle x(t) \rangle \equiv \sum_{n=0}^{N-1} \sum_{l_{n}} (n + l_{n}N) {\mathbb{p}_{n + l_{n}N}(t)}$, yielding drift velocity $v(t) \equiv {\rm d}\langle x(t) \rangle/{\rm d}t =\sum_{n=0}^{N-1} \sum_{l_{n}} (n + l_{n}N) {\dot{\mathbb{p}}_{n + l_{n}N}(t)}$.  Using the ME,  one obtains $v(t) = (\alpha -\beta) \sum_{n=0}^{N-1} \sum_{l_{n}} {\mathbb{p}_{n+l_{n}N}(t)} = (\alpha -\beta) \sum_{n=0}^{N-1} p_{n}(t)$. As $t \to \infty$,  one obtains $ v^{\rm st}_{\rm {drift}}= (\alpha -\beta) \sum_{n=0}^{N-1} p^{\rm{st}} = (\alpha -\beta) N p^{\rm{st}}$.  The equivalence of the NESS dynamics on the RC-backbone with that of $1d$ periodic system implies $ v^{\rm st}_{\rm {drift}}=(\alpha -\beta) N P^{\rm{st}}$ for RC, obtaining
\begin{eqnarray}
	v^{\rm st}_{\rm {drift}} = \frac{(\alpha -\beta)}{1 + \frac{1}{N}  \sum_{n=0}^{N-1} \Delta_{L_{n}}(s,r)|_{s \to 0} }.
	\label{eq:drift_vel}
\end{eqnarray}
The result~(\ref{eq:drift_vel}) is verified in Fig.~\ref{fig:drift_vel_numerics}(c) against numerical simulations for $N=200$,~$W=0.5$,  exponential $\mathcal{P}_{L}$ ($M=20$, $\xi=5$)~\cite{SM}. 

Note that $v^{\rm st}_{\rm drift}$ in Eq.~(\ref{eq:drift_vel}) gives the drift velocity for a given disorder realization.  As $N \to \infty$, the law of large numbers lets the sample average $(1/N)  \sum_{n=0}^{N-1}  \Delta_{L_{n}}(s,r)|_{s \to 0}$ in Eq.~(\ref{eq:drift_vel}) be replaced with expectation $\langle \Delta_{L}(s,r)|_{s \to 0} \rangle \equiv \sum_L \Delta_{L}(s,r)|_{s\to 0} \mathcal{P}_L $, when the latter is finite, as is the case with finite $M$.  Such a replacement makes the resulting expression independent of disorder realizations: $v^{\rm st}_{\rm drift} \to \overline{v^{\rm st}_{\rm {drift}}}$,  with overbar denoting the disorder-realization-independent answer. 

For exponential $\mathcal{P}_{L}$, one easily computes for $r=0$ the quantity $\langle \Delta_{L_{n}}(s,r=0)|_{s \to 0} \rangle$,  obtaining
\begin{eqnarray}
	\overline{v^{\rm st}_{\rm {drift}}} = \frac{(\alpha -\beta)}{e^{ 1/L(g)}-1} \left[  \frac{e^{1/L(g)} G_{M}\left(e^{ - 1/\xi} \right)}{G_{M} \left(e^{1/L(g) - 1/\xi}\right)}-1\right],
	\label{eq:v_drift_r_zero}
\end{eqnarray}  
with $G_{M}(y) \equiv ({1-y})/({1-y^{M+1}})$ and the bias-dependent length scale $L(g) \equiv 1/{\ln f} = [{\ln \left( ({1+g})/({1-g})\right)}]^{-1}$~\cite{white1984field}.  For $r=0$,  $\overline{P_{n,m}^{\rm st}}=f^m \overline{P^{\rm st}}$ implies that the net stationary-state probability-current due to biased-RW dynamics, $\overline{J_{(n,m-1) \to (n,m)}^{\rm {st,~RW}}} \equiv\alpha  \overline{P_{n, m-1}^{\rm {st}}} - \beta \overline{P_{n, m}^{\rm {st}}}$, is zero in the branches, which was a crucial assumption to derive Eq.~(\ref{eq:v_drift_r_zero}) in Ref.~\cite{white1984field} and that we show here to be exact. In contrast, for $r \neq 0$,  $\overline{J_{(n,m-1) \to (n,m)}^{{\rm {st,~RW}}}} > 0$, and the difference of the net stationary-state probability-current into and out of a site is balanced by an outgoing resetting current~\cite{SM}. 

For finite $N,~M$, the sample average in Eq.~(\ref{eq:drift_vel}) is finite, and so is $v^{\rm st}_{\rm drift}$; $N \to \infty$ at finite $M$, when $\langle \Delta_{L}|_{s \to 0} \rangle$ is always finite, too yields finite $v^{\rm st}_{\rm drift}$.  The opposite limit $M \to \infty$ at finite $N$ may render the sample average infinite,  yielding $v^{\rm st}_{\rm drift}=0$ for specific disorder realizations. A case of interest is considering limit $N \to \infty$ first,  when expectations replace sample averages, followed by $M \to \infty$, and asking: does the disorder-realization-independent $\overline{v^{\rm st}_{\rm drift}}$ become zero at any $g$? For $\overline{v^{\rm st}_{\rm drift}}$ to be zero, $\langle \Delta_{L}(s,r)|_{s \to 0} \rangle$ has to diverge. {Now, we have $\langle \Delta_{L}(s,r)|_{s \to 0} \rangle=\sum_L \mathcal{P}_L\Delta_{L}(s,r)|_{s \to 0}$, wherein, while $\mathcal{P}_L$ is always finite and is a decreasing function of $L$, the quantity $\Delta_{L}(s,r)|_{s \to 0}$ is an increasing function of $L$ with $\Delta_{L}(s,r)|_{s \to 0}$ becoming zero at $L = 0$. Consequently, the product $\mathcal{P}_L\Delta_{L}(s,r)|_{s \to 0}$ will be either (i) a monotonically increasing function of $L$ that diverges as $L \to \infty$, or, (ii) a monotonically decreasing function of $L$ that does not ever diverge at any $L$ and goes to zero as $L \to \infty$, or, (iii) a nonmonotonic function of $L$ that goes to zero at $L = 0$ and as $L \to \infty$, with a peak at a finite value $L^\star$ of $L$. Then, as $M \to \infty$, one has the quantity $\langle \Delta_{L}(s,r)|_{s \to 0} \rangle$ remaining finite in cases (ii) and (iii); in the case of (i), however, $\langle \Delta_{L}(s,r)|_{s \to 0} \rangle$ will be diverging, owing to the term $\Delta_{M}(s,r)|_{s \to 0} \mathcal{P}_{M}$ tending to infinity as $M \to \infty$. We thus conclude that divergence of $\langle \Delta_{L}|_{s \to 0} \rangle$ requires $ \lim_{M \to \infty}~\Delta_{M}|_{s \to 0} \mathcal{P}_{M} \to \infty$, where we have for brevity suppressed the dependence of $\Delta_L$ on $s$ and $r$.} If $n^*$ is a backbone site with attached branch length $M$,  $\Delta_{M}|_{s \to 0}  \mathcal{P}_{M}  = (1/\overline{ P^{\rm{st}}})\sum_{m'=1}^{M} \overline{P_{n^*,m'}^{\rm st}} \mathcal{P}_{M} $ diverges in the limit $M \to \infty$ if 
\begin{eqnarray}
	\lim_{M \to \infty} \left(\mathcal{R} \equiv \mathcal{P}_{M}\overline{P_{n^*,M}^{\rm st}}/\overline{P^{\rm st}}\right)\to \infty.
	\label{eq:trapping_condition}
\end{eqnarray}

Physically,  $\mathcal{R}$ represents the contribution, from those backbone sites with attached branch length equal to $M$ to the quantity $\langle \Delta_{L}|_{s \to 0} \rangle$, of the relative probability $\overline{P_{n^*,M}^{\rm st}}/\overline{P^{\rm st}}$ of walkers to be on the open end of the branch to that on the backbone.  Now,  $\overline{P_{n^*,M}^{\rm st}}$ being a probability can never diverge.  Then, a diverging $\mathcal{R}$ that is associated with a zero drift implies that the walkers are trapped at the open end of such branches, so that one has a vanishing probability of finding them on the backbone: $\overline{P^{\rm st}}=0$.  Such a trapping results when a walker that happens to be at the open end of a branch at any time has to move against the bias to get to the backbone.  Equation~(\ref{eq:trapping_condition}) thus gives the criterion to observe trapping and hence a vanishing $\overline{v^\mathrm{st}_\mathrm{drift}}$.

With no resetting,  using $\overline{P_{n^*,M}^\mathrm{st}}=f^{M} \overline{P^\mathrm{st}}$,  we get for exponential $\mathcal{P}_{L}$ that $
\mathcal{R} \sim \exp \left[ M \left({1}/{L(g)} - {1}/{\xi} \right)\right],$
involving two competing length scales $\xi$ and $L(g)$.
As $M\to\infty$, trapping requires that $L(g) < \xi$.  Trapping causes a vanishing $\overline{v^\mathrm{st}_\mathrm{drift}}$. Thus, $\overline{v^\mathrm{st}_{\mathrm{drift}}}$ crosses over from a finite value to zero at $g=g_c$ satisfying $L(g_c)=\xi$. Our derived condition for trapping for exponential $\mathcal{P}_L$ was obtained in Ref.~\cite{white1984field} by analyzing $\overline{v_\mathrm{drift}^\mathrm{st}}$ in  Eq.~(\ref{eq:v_drift_r_zero}) as $M \to \infty$. We here go beyond Ref.~\cite{white1984field} in deriving the condition (\ref{eq:trapping_condition}) for trapping that is applicable to any distribution $\mathcal{P}_{L}$. For instance, for power-law $\mathcal{P}_{L}$ with $k>2$ so that $\langle f^{L_n+1}\rangle$ and hence, $\overline{P^\mathrm{st}}$ is finite,   $\mathcal{R} \sim \exp [ {M}/{L(g)} - k \ln {M}]$ diverges as $M \to \infty$ for any $0<g<1$,  implying $\overline{v^\mathrm{st}_\mathrm{drift}} = 0$ at any bias.

In the above backdrop, a pertinent question arises: what happens to trapping as one introduces infinitesimal resetting? Using $\mathcal{R}=(\mathcal{P}_{M}\Lambda_{M,M})/\Lambda_{0, M}$, exponential $\mathcal{P}_{L}$, and the limit $r \to 0$ yield for large $M$ the result~\cite{SM}: $\mathcal{ R} \sim \exp \left[ M \left({1}/{L(g)} - {1}/{\xi} \right)\right]  \exp \left[ - (r/(2Wg)) (f/(f-1)) f^{M} \right]$, in which the exponential involving $r$ gives the leading contribution in view of $f>1$.  Consequently, one has $\mathcal{R} \to 0$ as $M \to \infty$,  leading to a finite $\overline{v^\mathrm{st}_\mathrm{drift}}$ at any $g$.  The power-law $\mathcal{P}_{L}$ and $r \to 0$ yield for large $M$ that $\mathcal{ R} \sim \exp [ {M}/{L(g)} - k \ln {M}]  \exp \left[ - (r/(2Wg)) (f/(f-1)) f^{M} \right]$. Again,  it is because of the exponential involving $r$ that condition~(\ref{eq:trapping_condition}) is not satisfied, yielding a finite $\overline{v^\mathrm{st}_\mathrm{drift}}$ at any $g$. We thus see a dramatic consequence of resetting: while in its absence, on varying $g$, $\overline{v^\mathrm{st}_\mathrm{drift}}$ is zero for power-law $\mathcal{P}_L$ or shows a crossover from a finite value to zero for exponential $\mathcal{P}_L$,  it is \textit{always finite} in presence of resetting.

Finally, we study how $\overline{v^\mathrm{st}_\mathrm{drift}}$ changes on introducing infinitesimal resetting. Equation~(\ref{eq:drift_vel}) yields~\cite{SM}
\begin{eqnarray}
	\overline{v^\mathrm{st}_\mathrm{drift}} (r \to 0) - \overline{v^\mathrm{st}_\mathrm{drift}} (r = 0) = r \frac{ \alpha (\alpha -\beta)\langle b_{2} \rangle}{ \left(1 + \alpha \langle b_{1} \rangle \right)^{2} },
	\label{eq:drift_vel_r_to_zero_approx}
\end{eqnarray}
with $\langle b_{1} \rangle \equiv (2 W g)^{-1}\left( \langle f^{L_{n}} \rangle -1 \right)$,  and $ \langle b_{2} \rangle \equiv (4 W^{2} g^{2})^{-1} \Bigl[ (\langle f^{2 L_{n} +1 }\rangle -1)/{(f-1)}$ $ -  2 \langle L_{n} f^{L_{n}} \rangle  - \langle f^{L_{n}} \rangle \Bigr] $.  For any $\mathcal{P}_L$, the rhs is non-zero at any $g$, implying finite $\overline{v^\mathrm{st}_\mathrm{drift}}$ and no trapping on turning on resetting.  This is consistent with our earlier discussion on trapping condition not satisfied with resetting. A finite mean time $1/r$ between successive resets guarantees that a walker that is trapped at the open end of a long branch in absence of resetting can in its presence get to the backbone instantaneously through the now-allowed direct jump, thus avoiding trapping.

An interesting follow-up involves extending our analysis to a many-particle setup with exclusion interaction~\cite{ramaswamy1987transport} and employing reset-setups using optical tweezers \cite{besga2020optimal, tal2020experimental} to study RC-dynamics.  

MS thanks HPCE, IIT Madras for providing high performance computing facilities in AQUA cluster. SG acknowledges support from the Science and Engineering Research Board (SERB), India under SERB-MATRICS  Grant MTR/2019/000560 and SERB-CRG Grant CRG/2020/000596. SG also thanks ICTP, Trieste, Italy, for support under its Regular Associateship scheme.

\newcommand{\newblock}{}
\bibliographystyle{unsrtnat}
\bibliographystyle{iopart-num}
\bibliography{References_mrinal}

\begin{thebibliography}{70}
\providecommand{\natexlab}[1]{#1}
\providecommand{\url}[1]{\texttt{#1}}
\expandafter\ifx\csname urlstyle\endcsname\relax
  \providecommand{\doi}[1]{doi: #1}\else
  \providecommand{\doi}{doi: \begingroup \urlstyle{rm}\Url}\fi

\bibitem[de~Gennes et~al.(1976)]{de1976percolation}
Pierre~Gilles de~Gennes et~al.
\newblock La percolation: un concept unificateur.
\newblock \emph{La recherche}, 7\penalty0 (72):\penalty0 919--927, 1976.

\bibitem[Barma and Dhar(1983)]{barma1983directed}
Mustansir Barma and Deepak Dhar.
\newblock Directed diffusion in a percolation network.
\newblock \emph{Journal of Physics C: Solid State Physics}, 16\penalty0
  (8):\penalty0 1451, 1983.

\bibitem[White and Barma(1984)]{white1984field}
Steven~R White and Mustansir Barma.
\newblock Field-induced drift and trapping in percolation networks.
\newblock \emph{Journal of Physics A: Mathematical and General}, 17\penalty0
  (15):\penalty0 2995, 1984.

\bibitem[Dhar(1984)]{dhar1984diffusion}
Deepak Dhar.
\newblock Diffusion and drift on percolation networks in an external field.
\newblock \emph{Journal of Physics A: Mathematical and General}, 17\penalty0
  (5):\penalty0 L257, 1984.

\bibitem[Bunde et~al.(1986)Bunde, Havlin, Stanley, Trus, and
  Weiss]{bunde1986diffusion}
A~Bunde, S~Havlin, HE~Stanley, B~Trus, and GH~Weiss.
\newblock Diffusion in random structures with a topological bias.
\newblock \emph{Physical Review B}, 34\penalty0 (11):\penalty0 8129, 1986.

\bibitem[Goldhirsch and Gefen(1987)]{goldhirsch1987biased}
I~Goldhirsch and Y~Gefen.
\newblock Biased random walk on networks.
\newblock \emph{Physical Review A}, 35\penalty0 (3):\penalty0 1317, 1987.

\bibitem[Havlin and Ben-Avraham(1987)]{havlin1987diffusion}
Shlomo Havlin and Daniel Ben-Avraham.
\newblock Diffusion in disordered media.
\newblock \emph{Advances in physics}, 36\penalty0 (6):\penalty0 695--798, 1987.

\bibitem[Havlin et~al.(1987)Havlin, Kiefer, and Weiss]{havlin1987anomalous}
Shlomo Havlin, James~E Kiefer, and George~H Weiss.
\newblock Anomalous diffusion on a random comblike structure.
\newblock \emph{Physical Review A}, 36\penalty0 (3):\penalty0 1403, 1987.

\bibitem[Aslangul et~al.(1994)Aslangul, Pottier, and
  Chvosta]{aslangul1994analytic}
C~Aslangul, N~Pottier, and P~Chvosta.
\newblock Analytic study of a model of diffusion on a random comblike
  structure.
\newblock \emph{Physica A: Statistical Mechanics and its Applications},
  203\penalty0 (3-4):\penalty0 533--565, 1994.

\bibitem[Balakrishnan and Van~den Broeck(1995)]{balakrishnan1995transport}
V~Balakrishnan and C~Van~den Broeck.
\newblock Transport properties on a random comb.
\newblock \emph{Physica A: Statistical Mechanics and its Applications},
  217\penalty0 (1-2):\penalty0 1--21, 1995.

\bibitem[Pottier(1995)]{pottier1995diffusion}
No{\"e}lle Pottier.
\newblock Diffusion on random comblike structures: field-induced trapping
  effects.
\newblock \emph{Physica A: Statistical Mechanics and its Applications},
  216\penalty0 (1-2):\penalty0 1--19, 1995.

\bibitem[Mitran et~al.(2013)Mitran, Melchert, and Hartmann]{mitran2013biased}
TL~Mitran, O~Melchert, and AK~Hartmann.
\newblock Biased and greedy random walks on two-dimensional lattices with
  quenched randomness: The greedy ant within a disordered environment.
\newblock \emph{Physical Review E}, 88\penalty0 (6):\penalty0 062101, 2013.

\bibitem[Yuste et~al.(2016)Yuste, Abad, and Baumgaertner]{yuste2016anomalous}
Santos~B Yuste, Enrique Abad, and Artur Baumgaertner.
\newblock Anomalous diffusion and dynamics of fluorescence recovery after
  photobleaching in the random-comb model.
\newblock \emph{Physical Review E}, 94\penalty0 (1):\penalty0 012118, 2016.

\bibitem[Demaerel and Maes(2018)]{demaerel2018death}
Thibaut Demaerel and Christian Maes.
\newblock Death and resurrection of a current by disorder, interaction or
  periodic driving.
\newblock \emph{Journal of Statistical Physics}, 173\penalty0 (1):\penalty0
  99--119, 2018.

\bibitem[Hart et~al.(2020)Hart, De~Tomasi, and Castelnovo]{hart2020compact}
Oliver Hart, Giuseppe De~Tomasi, and Claudio Castelnovo.
\newblock From compact localized states to many-body scars in the random
  quantum comb.
\newblock \emph{Physical Review Research}, 2\penalty0 (4):\penalty0 043267,
  2020.

\bibitem[Kotak and Barma(2022)]{kotak2022bias}
Jesal~D Kotak and Mustansir Barma.
\newblock Bias induced drift and trapping on random combs and the bethe
  lattice: Fluctuation regime and first order phase transitions.
\newblock \emph{Physica A: Statistical Mechanics and its Applications}, page
  127311, 2022.

\bibitem[Stauffer(1979)]{stauffer1979scaling}
Dietrich Stauffer.
\newblock Scaling theory of percolation clusters.
\newblock \emph{Physics reports}, 54\penalty0 (1):\penalty0 1--74, 1979.

\bibitem[Rammal and Toulouse(1983)]{rammal1983random}
Rammal Rammal and G{\'e}rard Toulouse.
\newblock Random walks on fractal structures and percolation clusters.
\newblock \emph{Journal de Physique Lettres}, 44\penalty0 (1):\penalty0 13--22,
  1983.

\bibitem[Sahimi(1993)]{sahimi1993flow}
Muhammad Sahimi.
\newblock Flow phenomena in rocks: from continuum models to fractals,
  percolation, cellular automata, and simulated annealing.
\newblock \emph{Reviews of modern physics}, 65\penalty0 (4):\penalty0 1393,
  1993.

\bibitem[M{\'e}ndez and Iomin(2013)]{mendez2013comb}
Vicen{\c{c}} M{\'e}ndez and Alexander Iomin.
\newblock Comb-like models for transport along spiny dendrites.
\newblock \emph{Chaos, Solitons \& Fractals}, 53:\penalty0 46--51, 2013.

\bibitem[Cecchi and Magnasco(1996)]{cecchi1996negative}
Guillermo~A Cecchi and Marcelo~O Magnasco.
\newblock Negative resistance and rectification in brownian transport.
\newblock \emph{Physical review letters}, 76\penalty0 (11):\penalty0 1968,
  1996.

\bibitem[Iomin(2012)]{iomin2012superdiffusive}
Alexander Iomin.
\newblock Superdiffusive comb: Application to experimental observation of
  anomalous diffusion in one dimension.
\newblock \emph{Physical Review E}, 86\penalty0 (3):\penalty0 032101, 2012.

\bibitem[Agliari et~al.(2014)Agliari, Blumen, and Cassi]{agliari2014slow}
Elena Agliari, Alexander Blumen, and Davide Cassi.
\newblock Slow encounters of particle pairs in branched structures.
\newblock \emph{Physical Review E}, 89\penalty0 (5):\penalty0 052147, 2014.

\bibitem[B{\'e}nichou et~al.(2015)B{\'e}nichou, Illien, Oshanin, Sarracino, and
  Voituriez]{benichou2015diffusion}
O~B{\'e}nichou, P~Illien, G~Oshanin, A~Sarracino, and R~Voituriez.
\newblock Diffusion and subdiffusion of interacting particles on comblike
  structures.
\newblock \emph{Physical review letters}, 115\penalty0 (22):\penalty0 220601,
  2015.

\bibitem[Iomin(2006)]{iomin2006toy}
A~Iomin.
\newblock Toy model of fractional transport of cancer cells due to
  self-entrapping.
\newblock \emph{Physical Review E}, 73\penalty0 (6):\penalty0 061918, 2006.

\bibitem[Campos et~al.(2006)Campos, Fort, and M{\'e}ndez]{campos2006transport}
Daniel Campos, Joaquim Fort, and Vicen{\c{c}} M{\'e}ndez.
\newblock Transport on fractal river networks: Application to migration fronts.
\newblock \emph{Theoretical population biology}, 69\penalty0 (1):\penalty0
  88--93, 2006.

\bibitem[Evans and Majumdar(2011)]{r0}
Martin~R Evans and Satya~N Majumdar.
\newblock Diffusion with stochastic resetting.
\newblock \emph{Physical review letters}, 106\penalty0 (16):\penalty0 160601,
  2011.

\bibitem[Evans et~al.(2020)Evans, Majumdar, and Schehr]{evans2020stochastic}
Martin~R Evans, Satya~N Majumdar, and Gr{\'e}gory Schehr.
\newblock Stochastic resetting and applications.
\newblock \emph{Journal of Physics A: Mathematical and Theoretical},
  53\penalty0 (19):\penalty0 193001, 2020.

\bibitem[Pal(2015)]{r7}
Arnab Pal.
\newblock Diffusion in a potential landscape with stochastic resetting.
\newblock \emph{Physical Review E}, 91\penalty0 (1):\penalty0 012113, 2015.

\bibitem[Nagar and Gupta(2016)]{nagar2016diffusion}
Apoorva Nagar and Shamik Gupta.
\newblock Diffusion with stochastic resetting at power-law times.
\newblock \emph{Physical Review E}, 93\penalty0 (6):\penalty0 060102, 2016.

\bibitem[Majumdar and Oshanin(2018)]{majumdar2018spectral}
Satya~N Majumdar and Gleb Oshanin.
\newblock Spectral content of fractional brownian motion with stochastic reset.
\newblock \emph{Journal of Physics A: Mathematical and Theoretical},
  51\penalty0 (43):\penalty0 435001, 2018.

\bibitem[Den~Hollander et~al.(2019)Den~Hollander, Majumdar, Meylahn, and
  Touchette]{den2019properties}
Frank Den~Hollander, Satya~N Majumdar, Janusz~M Meylahn, and Hugo Touchette.
\newblock Properties of additive functionals of brownian motion with resetting.
\newblock \emph{Journal of Physics A: Mathematical and Theoretical},
  52\penalty0 (17):\penalty0 175001, 2019.

\bibitem[Chatterjee et~al.(2018)Chatterjee, Christou, and Schadschneider]{r17}
Abhinava Chatterjee, Christos Christou, and Andreas Schadschneider.
\newblock Diffusion with resetting inside a circle.
\newblock \emph{Physical Review E}, 97\penalty0 (6):\penalty0 062106, 2018.

\bibitem[Masoliver(2019)]{masoliver2019telegraphic}
Jaume Masoliver.
\newblock Telegraphic processes with stochastic resetting.
\newblock \emph{Physical Review E}, 99\penalty0 (1):\penalty0 012121, 2019.

\bibitem[Ray and Reuveni(2020)]{ray2020diffusion}
Somrita Ray and Shlomi Reuveni.
\newblock Diffusion with resetting in a logarithmic potential.
\newblock \emph{The Journal of chemical physics}, 152\penalty0 (23):\penalty0
  234110, 2020.

\bibitem[Montero and Villarroel(2016)]{montero2016directed}
Miquel Montero and Javier Villarroel.
\newblock Directed random walk with random restarts: The sisyphus random walk.
\newblock \emph{Physical Review E}, 94\penalty0 (3):\penalty0 032132, 2016.

\bibitem[M{\'e}ndez and Campos(2016)]{r20}
Vicen{\c{c}} M{\'e}ndez and Daniel Campos.
\newblock Characterization of stationary states in random walks with stochastic
  resetting.
\newblock \emph{Physical Review E}, 93\penalty0 (2):\penalty0 022106, 2016.

\bibitem[Kusmierz et~al.(2014{\natexlab{a}})Kusmierz, Majumdar, Sabhapandit,
  and Schehr]{kusmierz2014first}
Lukasz Kusmierz, Satya~N Majumdar, Sanjib Sabhapandit, and Gr{\'e}gory Schehr.
\newblock First order transition for the optimal search time of l{\'e}vy
  flights with resetting.
\newblock \emph{Physical review letters}, 113\penalty0 (22):\penalty0 220602,
  2014{\natexlab{a}}.

\bibitem[Belan(2018)]{belan2018restart}
Sergey Belan.
\newblock Restart could optimize the probability of success in a bernoulli
  trial.
\newblock \emph{Physical review letters}, 120\penalty0 (8):\penalty0 080601,
  2018.

\bibitem[Coghi and Harris(2020)]{coghi2020large}
Francesco Coghi and Rosemary~J Harris.
\newblock A large deviation perspective on ratio observables in reset
  processes: robustness of rate functions.
\newblock \emph{Journal of Statistical Physics}, 179\penalty0 (1):\penalty0
  131--154, 2020.

\bibitem[Kumar et~al.(2020)Kumar, Sadekar, and Basu]{kumar2020active}
Vijay Kumar, Onkar Sadekar, and Urna Basu.
\newblock Active brownian motion in two dimensions under stochastic resetting.
\newblock \emph{Physical Review E}, 102\penalty0 (5):\penalty0 052129, 2020.

\bibitem[Bressloff(2020)]{bressloff2020modeling}
Paul~C Bressloff.
\newblock Modeling active cellular transport as a directed search process with
  stochastic resetting and delays.
\newblock \emph{Journal of Physics A: Mathematical and Theoretical},
  53\penalty0 (35):\penalty0 355001, 2020.

\bibitem[Evans et~al.(2013)Evans, Majumdar, and Mallick]{evans2013optimal}
Martin~R Evans, Satya~N Majumdar, and Kirone Mallick.
\newblock Optimal diffusive search: nonequilibrium resetting versus equilibrium
  dynamics.
\newblock \emph{Journal of Physics A: Mathematical and Theoretical},
  46\penalty0 (18):\penalty0 185001, 2013.

\bibitem[Kusmierz et~al.(2014{\natexlab{b}})Kusmierz, Majumdar, Sabhapandit,
  and Schehr]{r3}
Lukasz Kusmierz, Satya~N Majumdar, Sanjib Sabhapandit, and Gr{\'e}gory Schehr.
\newblock First order transition for the optimal search time of l{\'e}vy
  flights with resetting.
\newblock \emph{Physical review letters}, 113\penalty0 (22):\penalty0 220602,
  2014{\natexlab{b}}.

\bibitem[Pal and Reuveni(2017)]{pal2017first}
Arnab Pal and Shlomi Reuveni.
\newblock First passage under restart.
\newblock \emph{Physical review letters}, 118\penalty0 (3):\penalty0 030603,
  2017.

\bibitem[Falc{\'o}n-Cort{\'e}s et~al.(2017)Falc{\'o}n-Cort{\'e}s, Boyer,
  Giuggioli, and Majumdar]{falcon2017localization}
Andrea Falc{\'o}n-Cort{\'e}s, Denis Boyer, Luca Giuggioli, and Satya~N
  Majumdar.
\newblock Localization transition induced by learning in random searches.
\newblock \emph{Physical review letters}, 119\penalty0 (14):\penalty0 140603,
  2017.

\bibitem[Chechkin and Sokolov(2018)]{chechkin2018random}
A~Chechkin and IM~Sokolov.
\newblock Random search with resetting: a unified renewal approach.
\newblock \emph{Physical review letters}, 121\penalty0 (5):\penalty0 050601,
  2018.

\bibitem[Bhat et~al.(2016)Bhat, De~Bacco, and Redner]{r12}
Uttam Bhat, Caterina De~Bacco, and S~Redner.
\newblock Stochastic search with poisson and deterministic resetting.
\newblock \emph{Journal of Statistical Mechanics: Theory and Experiment},
  2016\penalty0 (8):\penalty0 083401, 2016.

\bibitem[Ahmad et~al.(2019)Ahmad, Nayak, Bansal, Nandi, and Das]{r21}
Saeed Ahmad, Indrani Nayak, Ajay Bansal, Amitabha Nandi, and Dibyendu Das.
\newblock First passage of a particle in a potential under stochastic
  resetting: A vanishing transition of optimal resetting rate.
\newblock \emph{Physical Review E}, 99\penalty0 (2):\penalty0 022130, 2019.

\bibitem[Rold{\'a}n et~al.(2016)Rold{\'a}n, Lisica, S{\'a}nchez-Taltavull, and
  Grill]{r25}
{\'E}dgar Rold{\'a}n, Ana Lisica, Daniel S{\'a}nchez-Taltavull, and Stephan~W
  Grill.
\newblock Stochastic resetting in backtrack recovery by rna polymerases.
\newblock \emph{Physical Review E}, 93\penalty0 (6):\penalty0 062411, 2016.

\bibitem[Tucci et~al.(2020)Tucci, Gambassi, Gupta, and Rold{\'a}n]{r26}
Gennaro Tucci, Andrea Gambassi, Shamik Gupta, and {\'E}dgar Rold{\'a}n.
\newblock Controlling particle currents with evaporation and resetting from an
  interval.
\newblock \emph{Physical Review Research}, 2\penalty0 (4):\penalty0 043138,
  2020.

\bibitem[Reuveni(2016)]{reuveni2016optimal}
Shlomi Reuveni.
\newblock Optimal stochastic restart renders fluctuations in first passage
  times universal.
\newblock \emph{Physical review letters}, 116\penalty0 (17):\penalty0 170601,
  2016.

\bibitem[Boyer and Solis-Salas(2014)]{boyer2014random}
Denis Boyer and Citlali Solis-Salas.
\newblock Random walks with preferential relocations to places visited in the
  past and their application to biology.
\newblock \emph{Physical review letters}, 112\penalty0 (24):\penalty0 240601,
  2014.

\bibitem[Giuggioli et~al.(2019)Giuggioli, Gupta, and
  Chase]{giuggioli2019comparison}
Luca Giuggioli, Shamik Gupta, and Matt Chase.
\newblock Comparison of two models of tethered motion.
\newblock \emph{Journal of Physics A: Mathematical and Theoretical},
  52\penalty0 (7):\penalty0 075001, 2019.

\bibitem[Gupta et~al.(2014)Gupta, Majumdar, and Schehr]{r23}
Shamik Gupta, Satya~N Majumdar, and Gr{\'e}gory Schehr.
\newblock Fluctuating interfaces subject to stochastic resetting.
\newblock \emph{Physical review letters}, 112\penalty0 (22):\penalty0 220601,
  2014.

\bibitem[Gupta and Nagar(2016)]{r24}
Shamik Gupta and Apoorva Nagar.
\newblock Resetting of fluctuating interfaces at power-law times.
\newblock \emph{Journal of Physics A: Mathematical and Theoretical},
  49\penalty0 (44):\penalty0 445001, 2016.

\bibitem[Durang et~al.(2014)Durang, Henkel, and Park]{durang2014statistical}
Xavier Durang, Malte Henkel, and Hyunggyu Park.
\newblock The statistical mechanics of the coagulation--diffusion process with
  a stochastic reset.
\newblock \emph{Journal of Physics A: Mathematical and Theoretical},
  47\penalty0 (4):\penalty0 045002, 2014.

\bibitem[Magoni et~al.(2020)Magoni, Majumdar, and Schehr]{ising-resetting}
Matteo Magoni, Satya~N Majumdar, and Gr{\'e}gory Schehr.
\newblock Ising model with stochastic resetting.
\newblock \emph{Physical Review Research}, 2\penalty0 (3):\penalty0 033182,
  2020.

\bibitem[Sarkar and Gupta(2022)]{sarkar2022synchronization}
Mrinal Sarkar and Shamik Gupta.
\newblock Synchronization in the kuramoto model in presence of stochastic
  resetting.
\newblock \emph{arXiv preprint arXiv:2203.00339}, 2022.

\bibitem[Basu et~al.(2019)Basu, Kundu, and Pal]{basu2019symmetric}
Urna Basu, Anupam Kundu, and Arnab Pal.
\newblock Symmetric exclusion process under stochastic resetting.
\newblock \emph{Physical Review E}, 100\penalty0 (3):\penalty0 032136, 2019.

\bibitem[Karthika and Nagar(2020)]{karthika2020totally}
S~Karthika and A~Nagar.
\newblock Totally asymmetric simple exclusion process with resetting.
\newblock \emph{Journal of Physics A: Mathematical and Theoretical},
  53\penalty0 (11):\penalty0 115003, 2020.

\bibitem[Fuchs et~al.(2016)Fuchs, Goldt, and Seifert]{fuchs2016stochastic}
Jaco Fuchs, Sebastian Goldt, and Udo Seifert.
\newblock Stochastic thermodynamics of resetting.
\newblock \emph{EPL (Europhysics Letters)}, 113\penalty0 (6):\penalty0 60009,
  2016.

\bibitem[Mukherjee et~al.(2018)Mukherjee, Sengupta, and Majumdar]{r4}
B~Mukherjee, K~Sengupta, and Satya~N Majumdar.
\newblock Quantum dynamics with stochastic reset.
\newblock \emph{Physical Review B}, 98\penalty0 (10):\penalty0 104309, 2018.

\bibitem[SM()]{SM}
\emph{See Supplemental Material for several technical details on derivation of
  the equations.}

\bibitem[Domazetoski et~al.(2020)Domazetoski, Mas{\'o}-Puigdellosas, Sandev,
  M{\'e}ndez, Iomin, and Kocarev]{domazetoski2020stochastic}
Viktor Domazetoski, Axel Mas{\'o}-Puigdellosas, Trifce Sandev, Vicen{\c{c}}
  M{\'e}ndez, Alexander Iomin, and Ljupco Kocarev.
\newblock Stochastic resetting on comblike structures.
\newblock \emph{Physical Review Research}, 2\penalty0 (3):\penalty0 033027,
  2020.

\bibitem[Gonz{\'a}lez et~al.(2021)Gonz{\'a}lez, Riascos, and
  Boyer]{gonzalez2021diffusive}
Fernanda~H Gonz{\'a}lez, Alejandro~P Riascos, and Denis Boyer.
\newblock Diffusive transport on networks with stochastic resetting to multiple
  nodes.
\newblock \emph{Physical Review E}, 103\penalty0 (6):\penalty0 062126, 2021.

\bibitem[Singh et~al.(2021)Singh, Sandev, Iomin, and
  Metzler]{singh2021backbone}
RK~Singh, T~Sandev, A~Iomin, and R~Metzler.
\newblock Backbone diffusion and first-passage dynamics in a comb structure
  with confining branches under stochastic resetting.
\newblock \emph{Journal of Physics A: Mathematical and Theoretical},
  54\penalty0 (40):\penalty0 404006, 2021.

\bibitem[Ramaswamy and Barma(1987)]{ramaswamy1987transport}
Ramakrishna Ramaswamy and Mustansir Barma.
\newblock Transport in random networks in a field: interacting particles.
\newblock \emph{Journal of Physics A: Mathematical and General}, 20\penalty0
  (10):\penalty0 2973, 1987.

\bibitem[Besga et~al.(2020)Besga, Bovon, Petrosyan, Majumdar, and
  Ciliberto]{besga2020optimal}
Benjamin Besga, Alfred Bovon, Artyom Petrosyan, Satya~N Majumdar, and Sergio
  Ciliberto.
\newblock Optimal mean first-passage time for a brownian searcher subjected to
  resetting: experimental and theoretical results.
\newblock \emph{Physical Review Research}, 2\penalty0 (3):\penalty0 032029,
  2020.

\bibitem[Tal-Friedman et~al.(2020)Tal-Friedman, Pal, Sekhon, Reuveni, and
  Roichman]{tal2020experimental}
Ofir Tal-Friedman, Arnab Pal, Amandeep Sekhon, Shlomi Reuveni, and Yael
  Roichman.
\newblock Experimental realization of diffusion with stochastic resetting.
\newblock \emph{The journal of physical chemistry letters}, 11\penalty0
  (17):\penalty0 7350--7355, 2020.

\end{thebibliography}

\clearpage
\markboth{Supplemental Material}{Supplemental Material}

\begin{center}
  \textbf{\large Supplemental Material for ``Biased random walk on random networks in presence of stochastic resetting: Exact results''}\\[1em]
  Mrinal Sarkar$^{1,2}$ and Shamik Gupta$^3$\\
  \vspace{0.5em}
  {\small
  $^1$Department of Physics, Indian Institute of Technology Madras, Chennai 600036, India\\
  $^2$Institute for Theoretical Physics, University of Heidelberg, Philosophenweg 19, D-69120 Heidelberg, Germany\\
  $^3$Department of Theoretical Physics, Tata Institute of Fundamental Research, Homi Bhabha Road, Mumbai 400005, India
  }
\end{center}
\vspace{2em}


\renewcommand{\theequation}{S\arabic{equation}}
\renewcommand{\thetable}{S\arabic{table}}
\renewcommand{\thefigure}{S\arabic{figure}}
\setcounter{equation}{0}
\setcounter{table}{0}
\setcounter{figure}{0}
\newcommand{\implies}{\Rightarrow}

\section{Derivation of the recursion relation $\widetilde{P}_{n,m}(s) = \Gamma_{L_{n} -m +1}  \widetilde{P}_{n, m-1} (s)$ and Eq.\,(3) of the main text}

\label{SM:Soln_ME_Laplace}
We provide here the details on deriving the recursion relation between branch-site probabilities along with the definition of $\Gamma_\mathcal{M}$. Let us first exhibit the explicit form of the ME, Eq.~(2), given in the main text. For branch sites ($m \neq 0$), we have
\begin{eqnarray}
	       \fl \frac{\mathrm{d}}{{\mathrm{d}t}} P_{n,L_{n}}(t) 
		=  \alpha P_{n, L_{n}-1}(t) - (\beta + r) P_{n,L_{n}}(t),~~~  m = L_{n},~ \mathrm{and} ~\forall~ n,\\
		\fl \frac{\mathrm{d}}{\mathrm{d}t} P_{n,m}(t)
		=  \alpha P_{n, m-1}(t) + \beta P_{n,m+1}(t) - (\alpha + \beta + r) P_{n,m}(t),~~~0 < m < L_{n},~\mathrm{and} ~\forall~ n, \nonumber
\end{eqnarray}
and for backbone sites ($m=0$), we have
\begin{eqnarray}
	\fl \frac{\mathrm{d}}{\mathrm{d}t} P_{n,0}(t) 
		=  \alpha P_{n-1, 0}(t) + \beta P_{n+1,0}(t) + \beta P_{n,1}(t) - (2\alpha + \beta + r) P_{n,0}(t) \nonumber \\
		+ r \sum_{m'=0}^{L_{n}} P_{n,m'}(t),~~~ 0 < n < N-1, \nonumber \\
	\fl \frac{\mathrm{d}}{\mathrm{d}t} P_{0,0}(t)
		=  \alpha P_{N-1, 0}(t) + \beta P_{1,0}(t) + \beta P_{0,1}(t) - (2\alpha + \beta + r) P_{0,0}(t) \nonumber \\
		+ r \sum_{m'=0}^{L_{0}} P_{0,m'}(t),~~~n=0, \label{eq:ME_E_SM} \\
	\fl \frac{\mathrm{d}}{\mathrm{d}t} P_{N-1,0}(t)
		=  \alpha P_{N-2, 0}(t) + \beta P_{0,0}(t) + \beta P_{N-1,1}(t) - (2\alpha + \beta + r) P_{N-1,0}(t) \nonumber \\
		+ r \sum_{m'=0}^{L_{N-1}} P_{N-1,m'}(t),~~~n=N-1. \nonumber	
\end{eqnarray}

On applying the Laplace transformation (LT), $\widetilde{P}_{n,m}(s) \equiv \int_{0}^{\infty} \mathrm{ d}t~e^{-st} P_{n,m}(t)$, to the reflecting end of the $n$-th branch ($m=L_{n}$) first yields
\begin{eqnarray}
\fl	s \widetilde{P}_{n,L_{n}}(s)=\alpha \widetilde{P}_{n, L_{n}-1}(s) - (\beta + r) \widetilde{P}_{n,L_{n}}(s)  \implies 
	\widetilde{P}_{n, L_{n}-1}(s) = \left (\frac{s + \beta + r}{\alpha} \right )\widetilde{P}_{n,L_{n}}(s).
	\label{eq:ME_L_A}
\end{eqnarray}
Similarly, applying LT to other branch sites, $0<m<L_{n}$, one obtains the following relation:
\begin{eqnarray}
	(s+\alpha + \beta + r)\widetilde{P}_{n,m}(s)
	=  \alpha \widetilde{P}_{n, m-1}(s) + \beta \widetilde{P}_{n,m+1}(s).
	\label{eq:ME_L_B}
\end{eqnarray}
Take $m=L_{n} -1$ and substitute Eq.~(\ref{eq:ME_L_A}) in Eq.~(\ref{eq:ME_L_B}). This yields
\begin{eqnarray}
	\widetilde{P}_{n, L_{n}-1}(s) &&= \frac{1}{ \frac{s + \alpha+ \beta + r}{\alpha} - \frac{\beta}{\alpha} \frac{1}{\left (\frac{s + \beta + r}{\alpha} \right )}} \widetilde{P}_{n, L_{n}-2}(s),
	\label{eq:ME_L_step1}
\end{eqnarray}
which relates the LT-transformed probabilities on the branch sites at a distance $1$ and $2$ units from the reflecting end of the branch.  Equation~(\ref{eq:ME_L_B}), on further taking $m=L_{n} -2$ and using Eq.~(\ref{eq:ME_L_step1}), yields
\begin{eqnarray}
	\widetilde{P}_{n, L_{n}-2}(s) &&= \frac{1}{ \frac{s + \alpha+ \beta + r}{\alpha} - \frac{\beta}{\alpha}     \frac{1}{\left( \frac{s + \alpha+ \beta + r}{\alpha} - \frac{\beta}{\alpha} \frac{1}{\left (\frac{s + \beta + r}{\alpha} \right )} \right)} } \widetilde{P}_{n, L_{n}-3}(s). 
	\label{eq:ME_L_step2}
\end{eqnarray}
This relates the LT-transformed probabilities on the branch sites at a distance $2$ and $3$ units from the reflecting end of the branch. Substituting this way for $m$ in Eq.~(\ref{eq:ME_L_B}) successively, we obtain a relationship between LT-transformed probabilities on any two consecutive branch sites. We thus introduce a quantity $\Gamma_\mathcal{M}$ with $\mathcal{M} = 1,2,\cdots,L_{n}$ that relates the LT-transformed probabilities on two consecutive branch sites at distance $\mathcal{M}-1$ and $\mathcal{M}$ from the reflecting end. It is defined as
\begin{eqnarray}
	\Gamma_\mathcal{M}(s, \alpha, \beta, r) \equiv \frac{1}{ \frac{s + \alpha + \beta + r}{\alpha} - \frac{\beta}{\alpha}     \frac{1}{  \frac{s + \alpha+ \beta + r}{\alpha} - \frac{\beta}{\alpha}     \frac{1}{ \ddots \frac{s + \alpha+ \beta + r}{\alpha} - \frac{\beta}{\alpha} \frac{1}{\frac{s + \beta + r}{\alpha} } } }},
	\label{eq:Lambda_m_def_SM}
\end{eqnarray}
a finite continued fraction with total number of terms in the denominator being $\mathcal{M}$. Any two consecutive branch-site probabilities are thus related by
\begin{eqnarray}
	\widetilde{P}_{n,m}(s) = \Gamma_{L_{n} -m +1} \widetilde{P}_{n, m-1} (s),~~~m=1,2,3,\dots, L_{n}.
	\label{eq:P_nm_recursion_rel_SM}
\end{eqnarray}
The recursion relation (\ref{eq:P_nm_recursion_rel_SM}) along with Eq.~(\ref{eq:Lambda_m_def_SM}) are provided in the main text.

\section{Calculation of ${\Delta}_{L_{n}}$ of the main text}
We calculate here an explicit expression of the quantity $\Delta_{L_{n}}$ defined in the main text as
\begin{eqnarray}
\fl	\Delta_{L_{n}} \widetilde{P}_{n,0}(s)
	=\sum_{m'=1}^{L_{n}} f^{m'/2} \frac{\sinh (L_{n} -m' +1) \theta - \sqrt{f} \sinh (L_{n} -m') \theta}{\sinh (L_{n} + 1) \theta - \sqrt{f} \sinh (L_{n}) \theta} \widetilde{P}_{n, 0} (s).
	\label{eq:Delta_n_def_SM}
\end{eqnarray}
One can easily perform the geometric sums to show that
\begin{eqnarray}
\fl \sum_{m'=1}^{L_{n}} f^{m'/2} \sinh (L_{n} -m' +1) \theta \nonumber \\
\fl = \frac{\sqrt{f}}{2 \sqrt{f} \cosh \theta -(1+f)} \left[ \sqrt{f} \sinh (L_{n} +1) \theta - (\sqrt{f})^{L_{n} + 1} \sinh \theta - \sinh (L_{n} \theta)\right],&\mathrm{and} \nonumber\\
\fl \sum_{m'=1}^{L_{n}} f^{m'/2} \sinh (L_{n} -m') \theta \label{eq:Delta_n_inter_SM} \\
\fl	 =\frac{\sqrt{f}}{2 \sqrt{f} \cosh \theta -(1+f)} \left[ \sqrt{f} \sinh L_{n} \theta - (\sqrt{f})^{L_{n}} \sinh \theta - \sinh (L_{n}-1) \theta \right].  \nonumber 
\end{eqnarray}
Substituting Eq.~(\ref{eq:Delta_n_inter_SM}) in Eq.~(\ref{eq:Delta_n_def_SM}), and by noticing that ${\sqrt{f}}/\left(2 \sqrt{f} \cosh \theta -(1+f)\right) = {\sqrt{\alpha \beta}}/{(s+r)}$, we finally obtain
\begin{eqnarray}
\fl 	\Delta_{L_{n}} = \frac{\sqrt{\alpha \beta}}{s+r} \left[ \sqrt{f} - \frac{\sinh L_{n} \theta - \sqrt{f} \sinh (L_{n} -1) \theta}{\sinh (L_{n}+1) \theta - \sqrt{f} \sinh L_{n}  \theta} \right]=\frac{ \beta}{s+r} \left( f - \Gamma_{L_{n}} \right),
	\label{eq:Delta_n_closed_form}
\end{eqnarray}
which is provided in the main text.

\section{Probability current in the branches in the stationary state}
Here we compute explicitly the net probability-current due to biased-RW dynamics between any two consecutive branch sites in the NESS. To this end, consider a link between $(m-1)$-th and $m$-th sites of the $n$-th branch. In absence of resetting ($r=0$), the probability current will be solely due to biased-RW, and the net current is given by
\begin{eqnarray}
\fl \overline{J_{(n,m-1) \to (n,m)}^\mathrm{{st,~RW}}}  \equiv \alpha  \overline{P_{n, m-1}^\mathrm{{st}}} - \beta \overline{P_{n, m}^\mathrm{{st}}},  \nonumber\\
 =\beta \left[ f \overline{P_{n, m-1}^\mathrm{{st}}} -  \overline{P_{n, m}^\mathrm{{st}}} \right]  
=\beta \left[ f^m - f^m \right] \overline{P^\mathrm{{st}}}=0,
\end{eqnarray}
where we have used $\overline{P_{n,m}^\mathrm{st}}=f^m \overline{P^\mathrm{st}}$, as obtained in the main text. Hence, in absence of resetting, there is no net probability-current in the branches in the NESS.

In the presence of resetting, the net probability-current due to biased-RW dynamics is computed as follows:
\begin{eqnarray}
\fl \overline{J_{(n,m-1) \to (n,m)}^\mathrm{{st,~RW}}} = \alpha  \overline{P_{n, m-1}^\mathrm{{st}}} - \beta \overline{P_{n, m}^\mathrm{{st}}} =\beta \left[ f \overline{P_{n, m-1}^\mathrm{{st}}} -  \overline{P_{n, m}^\mathrm{{st}}} \right],  \nonumber\\
 =\beta \left[ f \frac{\overline{P_{n, m-1}^\mathrm{{st}}}}{\overline{P_{n, m}^\mathrm{{st}}}} - 1 \right] \overline{P_{n, m}^\mathrm{{st}}}= \beta \left[ f \frac{\Lambda_{m-1, L_{n}}}{\Lambda_{m, L_n}} - 1 \right] \overline{P_{n, m}^\mathrm{{st}}},
\label{eq:diffusion_current_1}
\end{eqnarray} 
where $\Lambda_{m, L_{n}} = (f^{m/2}/2) [  \lambda^{L_{n}-m} \left( \lambda - \sqrt{f} \right) -  \lambda^{-L_{n}+m} \left( 1/\lambda - \sqrt{f} \right) ]$, as defined in the main text. To have an estimate of the net probability-current, let us consider the case of an  infinitesimal resetting rate. In the limit $r \to 0$, one can easily show that 
\begin{eqnarray}
\fl	\lambda = \frac{2W + r}{2W \sqrt{1-g^{2}}}  \left[  1+ \sqrt{1- \frac{4 W^{2} (1 -g^{2})}{(2W +r)^{2}}} \right] = \sqrt{f} \left[ 1 + \left( \frac{r}{2Wg} -  \frac{1}{f-1} \frac{r^{2}}{4 W^{2} g^{2}} \right)\right],
\label{eq:lambda_approx_SM}
\end{eqnarray}
keeping terms upto second order of $r$. On substituting Eq.~(\ref{eq:lambda_approx_SM}) in the expressions of $\Lambda_{m-1, L_{n}}$ and $\Lambda_{m, L_{n}}$ and keeping terms upto first order of $r$ yield
\begin{eqnarray}
\fl	\frac{\Lambda_{m-1, L_{n}}}{\Lambda_{m, L_n}} = \frac{1}{\sqrt{f}}\frac{\lambda^{L_{n}-m+1} \left( \lambda - \sqrt{f} \right) -  \lambda^{-L_{n}+m-1} \left( 1/\lambda - \sqrt{f} \right)}{\lambda^{L_{n}-m} \left( \lambda - \sqrt{f} \right) -  \lambda^{-L_{n}+m} \left( 1/\lambda - \sqrt{f} \right)}, \nonumber\\
= \frac{1}{f} \left[ 1 + \frac{r}{2Wg} \left( f^{L_n -m +1} -1\right)\right]. 
	\label{eq:diff_current_coeff_ratio}
\end{eqnarray}
On substituting Eq.~(\ref{eq:diff_current_coeff_ratio}) in Eq.~(\ref{eq:diffusion_current_1}), one obtains finally
\begin{eqnarray}
\fl \overline{J_{(n,m-1) \to (n,m)}^\mathrm{{st,~RW}}} = r \frac{f^{L_n -m +1} -1}{f-1} \overline{P_{n, m}^\mathrm{{st}}} \qquad > 0 \qquad (\mathrm{for}~\overline{P_{n, m}^\mathrm{{st}}} \neq 0 ).
\label{eq:diffusion_current_2}	
\end{eqnarray} 
Equation~(\ref{eq:diffusion_current_2}) thus implies that there is a net probability-current (due to biased-RW dynamics) in the branches along the direction of the bias. Consider now three consecutive branch-sites, say, $(m-1)$-th, $m$-th and $(m+1)$-th, and compute the incoming and outgoing net probability-currents at the $m$-th branch site, i.e., $\overline{J_{(n,m-1) \to (n,m)}^\mathrm{{st,~RW}}}$ and $\overline{J_{(n,m) \to (n,m+1)}^\mathrm{{st,~RW}}}$, respectively. The difference between these two probability-currents, on using Eq.~(\ref{eq:diffusion_current_2}) and Eq.~(\ref{eq:diff_current_coeff_ratio}), and on further simplification, reads as
\begin{eqnarray}
\fl \overline{J_{(n,m-1) \to (n,m)}^\mathrm{{st,~RW}}} -  \overline{J_{(n,m) \to (n,m+1)}^\mathrm{{st,~RW}}} \nonumber \\
= \frac{r}{f-1} \left[ \left( f^{L_n -m +1} -1 \right)  \overline{P_{n, m}^\mathrm{{st}}} - \left( f^{L_n -m} -1 \right)  \overline{P_{n, m+1}^\mathrm{{st}}} \right],\nonumber\\
 = \frac{r}{f-1} \left[ \left( f^{L_n -m +1} -1 \right) - \left( f^{L_n -m} -1 \right)  \frac{\Lambda_{m+1, L_{n}}}{\Lambda_{m, L_n}} \right] \overline{P_{n, m}^\mathrm{{st}}}
= r \overline{P_{n, m}^\mathrm{{st}}}. 
\label{eq:diffusion_current_3}	
\end{eqnarray} 
The rhs of Eq.~(\ref{eq:diffusion_current_3}) can be interpreted as outgoing resetting current against the direction of the bias at the $m$-th branch site. One may check the consistency of Eq.~(\ref{eq:diffusion_current_3}) by noting that Eq.~(\ref{eq:diffusion_current_3}), on using the definition of $\overline{J_{(n,m-1) \to (n,m)}^\mathrm{{st,~RW}}}$ $\left( \equiv \alpha  \overline{P_{n, m-1}^\mathrm{{st}}} - \beta \overline{P_{n, m}^\mathrm{{st}}}\right)$, recovers the stationary-state ME for the $m$-th branch-site. In passing, note that although we have considered the limit $N\to \infty$ in the above derivation of net probability current, the same result holds true for finite $N$ too.

\section{Derivation of $\mathcal{R}$ explicitly in the presence of an infinitesimal resetting rate}
We derive here the explicit expression of $\mathcal{R}$ in the presence of an infinitesimal resetting rate. We have in the limit $N \to \infty$,
\begin{eqnarray}
	\frac{\overline{P_{n,L_{n}}^{\rm{st}}}}{\overline{P^{\rm{st}}}} = \frac{\Lambda_{L_n, L_n}}{\Lambda_{0, L_n}} = \frac{ f^{L_{n}/2}\left (\lambda - {1}/{\lambda} \right )}{  \lambda^{L_{n}} \left( \lambda - \sqrt{f} \right) -  \lambda^{-L_{n}} \left( 1/\lambda- \sqrt{f} \right) }.
	\label{eq:trapping_derivation_1_SM}
\end{eqnarray}
In the limit $r \to 0$, Eq.~(\ref{eq:lambda_approx_SM}) yields, $\lambda = \sqrt{f} \left[ 1 + \left( {r}/{(2Wg)} -  ({1}/{(f-1)}) \left( r^{2}/{(4 W^{2} g^{2})} \right) \right)\right]$. On substituting the value of $\lambda$ in the limit considered in Eq.~(\ref{eq:trapping_derivation_1_SM}) and keeping terms upto first order of $r$, a straightforward calculation yields
\begin{equation}
	\frac{\overline{P_{n,L_{n}}^{\rm{st}}}}{\overline{P^{\rm{st}}}} = \frac{\Lambda_{L_n, L_{n}}}{\Lambda_{0, L_{n}}} =  f^{L_{n}} \left[ 1 - \frac{r}{2Wg} \left\{ \frac{f^{L_n +1} -1}{f-1} - (L_n + 1)\right\}\right].
	\label{eq:trapping_derivation_0_SM}
\end{equation}
Note that one could also arrive at Eq.~(\ref{eq:trapping_derivation_0_SM}) using Eq.~(\ref{eq:diff_current_coeff_ratio}) recursively. We must remember that Eq.~(\ref{eq:trapping_derivation_0_SM}) is valid in the limit $r \to 0$ such that $(r/(2Wg)) \left\{ \left( {f^{L_n +1} -1} \right)/(f-1) - (L_n + 1)\right\} \ll 1$. Moreover, since $f >1$, for large $L_n$, the term inside the braces in Eq.~(\ref{eq:trapping_derivation_0_SM}) can be approximated by $(f/(f-1)) f^{L_n} $. Thus, we may write finally:
\begin{equation}
	\frac{\mathcal{P}_{L_n}\overline{P_{n,L_{n}}^{\rm{st}}}}{\overline{P^{\rm{st}}}}=\frac{\mathcal{P}_{L_n}\Lambda_{L_{n}, L_n}}{\Lambda_{0, L_n}} \approx \mathcal{P}_{L_n} f^{L_{n}} \exp \left( - \frac{r}{2Wg} \frac{f}{f-1} f^{L_{n}} \right).
	\label{eq:trapping_derivation_2_SM}
\end{equation}
For an exponential or power-law $\mathcal{P}_L$ with $L_n=M$, one obtains from Eq.~(\ref{eq:trapping_derivation_2_SM}) the corresponding explicit expression of $\mathcal{R}$ provided in the main text.

\section{Derivation of Eq.~(15) of the main text}
Here we will provide the derivation to obtain the behavior of $\overline{v^\mathrm{st}_\mathrm{{drift}}}$ on introducing an infinitesimal resetting in the dynamics. From Eqs.~(12) and (11) of the main text, we have for $r \neq 0$ and in the limit $N \to \infty$,
\begin{equation}
	\overline{v^\mathrm{st}_\mathrm{{drift}}} = \frac{(\alpha -\beta)}{ (1 + f \beta/r)  - (\beta/r) \langle \Gamma_{L_{n}}(s,r>0)|_{s \to 0} \rangle }, 
	\label{eq:drift_vel_SM}	
\end{equation}
where $ \langle \Gamma_{L_{n}}(s,r>0)|_{s \to 0} \rangle$ is defined in the main text. Note that there is a prefactor $1/r$ with $ \langle \Gamma_{L_{n}}(s,r>0)|_{s \to 0} \rangle$ in Eq.~(\ref{eq:drift_vel_SM}). We will thus study the behavior of $ \langle \Gamma_{L_{n}}(s,r>0)|_{s \to 0} \rangle$ in the limit $r \to 0$ keeping terms upto second order of $r$. In this limit, $\lambda$ reduces to, $\lambda = \sqrt{f} \left[ 1 + \left( {r}/{(2Wg)} -  ({1}/{(f-1)}) \left( r^{2}/{(4 W^{2} g^{2})} \right) \right)\right]$ [see Eq.~(\ref{eq:lambda_approx_SM})], which we substitute in the expression of $\Gamma_{L_{n}}(s,r>0)|_{s \to 0}$ and simplify to obtain
\begin{eqnarray}
	\Gamma_{L_{n}}|_{s \to 0} =  \frac{\Lambda_{1, L_{n}}}{\Lambda_{0, L_{n}}}=\sqrt{f} \frac{\lambda^{L_{n}-1} \left( \lambda - \sqrt{f} \right) -  \lambda^{-L_{n}+1}\left( 1/\lambda- \sqrt{f} \right)}{\lambda^{L_{n}} \left( \lambda - \sqrt{f} \right) -  \lambda^{-L_{n}}\left( {1}/{\lambda}- \sqrt{f} \right)}, \nonumber\\
	 = f (1 - b_{1} r + b_{2} r^{2}), 
	\label{eq:omega_st_2_SM}
\end{eqnarray}
where \begin{eqnarray}
	b_{1} =  \frac{f^{L_{n}} -1}{2 W g},\,\, \mathrm{and} \,\,b_{2} = \frac{1}{4 W^{2} g^{2}} \left[ \frac{f^{2 L_{n} +1 } -1}{f-1} -  \left( 2 L_{n} +1 \right) f^{L_{n}}\right].
	\label{eq:b1_b2_SM}
\end{eqnarray}
Note that both the coefficients $b_{1}$ and $b_{2}$ are positive for $L_{n} > 0$, whereas they vanish for $L_{n} = 0$.

On substituting Eq.~(\ref{eq:omega_st_2_SM}) in Eq.~(\ref{eq:drift_vel_SM}) yields
\begin{eqnarray}
\fl	\overline{v^\mathrm{st}_\mathrm{{drift}}} (r \to 0) =  \frac{(\alpha -\beta)}{ 1  - \alpha (\langle b_1 \rangle  - \langle b_2 \rangle r)} = \frac{(\alpha -\beta)}{ 1  - \alpha \langle b_1 \rangle } + r \frac{\alpha (\alpha -\beta) \langle b_2 \rangle }{ \left( 1  - \alpha \langle b_1 \rangle \right)^2} + \mathcal{O}(r^2).
	\label{eq:v_drift_r_to_zero_approx_SM_0}
\end{eqnarray}
Using Eq.~(\ref{eq:b1_b2_SM}), one can easily identify the first term ${(\alpha -\beta)}/\left( 1  - \alpha \langle b_1 \rangle \right) $ with the drift velocity in absence of resetting, $\overline{v^\mathrm{st}_\mathrm{{drift}}} (r = 0)$. We, therefore, obtain from Eq.~(\ref{eq:v_drift_r_to_zero_approx_SM_0}), keeping terms upto first order of $r$,
\begin{eqnarray}
	\overline{v^\mathrm{st}_\mathrm{{drift}}} (r \to 0) - \overline{v^\mathrm{st}_\mathrm {{drift}}} (r = 0) = r \frac{ \alpha (\alpha -\beta)\langle b_{2} \rangle}{ \left(1 + \alpha \langle b_{1} \rangle \right)^{2} },
	\label{eq:v_drift_r_to_zero_approx_SM}
\end{eqnarray}
with
\begin{eqnarray}
	\fl \langle b_{1} \rangle = \frac{1}{2 W g}\left( \langle f^{L_{n}} \rangle -1 \right), ~~\mathrm{and}\\
	\fl \langle b_{2} \rangle = \frac{1}{4 W^{2} g^{2}} \left[ \frac{ \langle f^{2 L_{n} +1 }\rangle -1}{f-1} -  2 \langle L_{n} f^{L_{n}} \rangle    - \langle f^{L_{n}} \rangle \right], \nonumber
\end{eqnarray}
which is provided in Eq.~(15) in the main text.

\section{Details of numerical simulation}
\label{SM:Num_sim}
We discuss here the details of numerical algorithm to simulate the model discussed in the main text, for given values of bias $g$, the parameter $W$ appearing in the hop rates,  the resetting rate $r$,  the number $N$ of backbone sites, and the branch-length cut-off $M$. We take the lattice spacing to be unity. In our simulations, we choose $W=0.5$, $N=200$ and an exponential $\mathcal{P}_L$ (Eq. (1) of the main text) with $\xi=5$, and $M=20$, unless stated otherwise.

However, one may choose any other distribution and implement the numerics. The $N$ branch lengths $L_{n}$'s that are quenched-disordered random variables are chosen independently from the exponential $\mathcal{P}_L$. The dynamics as detailed below proceeds for a given realization of the $L_n$'s, and we measure in numerics the values of macroscopic quantities such as drift velocity of the walkers (particles). Typical simulations involved initializing the dynamics at time $t=0$ with a particle at location $(n,m)=(0,0)$, and letting it perform dynamics in continuous time with chosen infinitesimal time interval ${\rm d}t$.  Given the position of the particle at time $t$, in the ensuing infinitesimal time interval $[t,t+{\rm d}t]$,  the position of the particle is updated as follows.  We draw a uniformly-distributed random number $R$ in $[0,1]$. If we find that $R < r{\rm d}t$, then the particle if on a branch site at time $t$ resets to the corresponding backbone site, while if the particle is already on a backbone site, it stays put. On the other hand,  if $R> r{\rm d}t$, then the particle performs biased random walk. (i) If the particle at time $t$ was on a branch site that is not the end site of the branch, then it decides to move along the branch with equal probability of $1/3$ in the direction of and opposite to the direction of the bias, while it decides to stay put with probability $1/3$.  The move in the direction of (respectively, opposite to the direction of) the bias is actually accepted with probability $(3/2)(1+g) {\rm d}t$ (respectively, with probability $(3/2)(1-g) {\rm d}t$).  (ii) If the particle at time $t$ was on the end site (reflecting end) of the branch, then it decides to move along the branch and opposite to the direction of the bias with probability $1/3$, while it decides to stay put with probability $2/3$.  The move is actually accepted with probability $(3/2)(1-g) {\rm d}t$.  (iii) If the particle was on a backbone site that has no branch attached to it, it moves along the backbone and in the direction of (respectively, opposite to the direction of) the bias with equal probability of $1/3$, while it stays put with probability $1/3$.  The moves in the direction of and opposite to the direction of the bias are accepted respectively with probabilities $(3/2)(1+g){\rm d}t$ and $(3/2)(1-g){\rm d}t$.  (iv) If the particle was on a backbone site that has a branch attached to it, it moves along the backbone and in the direction of (respectively, opposite to the direction of) the bias with equal probability of $1/3$, while it moves into the attached branch with probability $1/3$.  The moves in the direction of and opposite to the direction of the bias are accepted respectively with probabilities $(3/2)(1+g){\rm d}t$ and $(3/2)(1-g){\rm d}t$.  The move to the branch is accepted with probability $(3/2)(1+g){\rm d}t$. Averaging over independent dynamical realizations with one particle is tantamount to performing the dynamics with several particles performing independent dynamics. 

In numerics, we start with $\mathcal{N}=2000$ number of particles and keep evolving their dynamics following the updating rules mentioned above for a long time (so that the dynamics settles down to a stationary state), and measure the drift velocity at long times in the following way.

\begin{itemize}
	\item \textbf{Drift velocity:} At long times after initiating the dynamics, we start tracking individual particles for a long observation time $T$. Let us denote the particles by the index $i$ with $i=1, 2,\dots, \mathcal{N}$. Then we compute the velocity of individual particle ($v[i]$) on the backbone along the direction of the bias as follows:
	\begin{eqnarray}
		\fl v[i] = \frac{\mathrm{Net\,displacement\,of\,the}\, i \mathrm{-th\,particle\,on\,the\,backbone\,along\,the\,bias }}{\mathrm{Observation\,time} (T)}. 
	\end{eqnarray}
	The drift velocity is computed using the following
	\begin{eqnarray}
		v^{\rm st}_{\rm {drift}} = \frac{1}{\mathcal{N}} \sum_{i=1}^{\mathcal{N}} v[i].
	\end{eqnarray}
	
\end{itemize}

For a fixed value of resetting rate $r$, we repeat the above procedures for updating the dynamics and compute the drift velocity for various values of $g$. Finally, we repeat the whole study for various values of $r$.

\end{document}